\documentstyle[amsmath,amssymb,prd,tighten,preprint,epsfig,aps]
{revtex}

\newcommand{\cO}{\mathcal {O}}
\newcommand{\DD}{\not\!\!{D}}
\newcommand{\half}{\mbox{\small $\frac{1}{2}$}}
\newcommand{\third}{\mbox{\small $\frac{1}{3}$}}
\newcommand{\Dd}[1]{\overset{\leftrightarrow}{D}_{#1}}

\newlength{\ww}

\begin{document}
\draft
\settowidth{\ww}{HU-EP-00/50}
\preprint{\parbox{\ww}{
DESY 00-168
\\ HU-EP-00/50 
}}

\title{
A lattice calculation of the nucleon's spin-dependent structure function
$g_2$ revisited}
\author{M. G\"ockeler$^1$, R. Horsley$^{2,3}$, W. K\"urzinger$^{4,2}$, 
   H. Oelrich$^2$, D. Pleiter$^{4,2}$, P.E.L. Rakow$^1$, \\
   A. Sch\"afer$^1$, and G. Schierholz$^{2,5}$}
\address{
         $^1$Institut f\"ur Theoretische Physik, Universit\"at Regensburg,
                    D-93040 Regensburg,
                    Germany \\
        $^2$Deutsches Elektronen-Synchrotron DESY,
             John von Neumann-Institut f\"ur Computing NIC, \\
                    D-15735 Zeuthen, Germany \\
        $^3$Institut f\"ur Physik, Humboldt-Universit\"at zu Berlin,
                    D-10115 Berlin, Germany \\
        $^4$Institut f\"ur Theoretische Physik, Freie Universit\"at Berlin,
                    D-14195 Berlin, Germany \\
        $^5$Deutsches Elektronen-Synchrotron DESY,
                    D-22603 Hamburg, Germany}

\maketitle

\begin{abstract}
Our previous calculation of the spin--dependent structure function $g_2$ 
is revisited. The interest in this structure function is to a great extent
motivated by the fact that it receives contributions from twist--two as 
well as from twist--three operators already in leading order of $1/Q^2$
thus offering the unique possibility of directly assessing higher--twist
effects.
In our former  calculation the lattice operators were renormalized
perturbatively and mixing with lower--dimensional operators was ignored.
However, the twist--three operator which gives rise to the matrix element
$d_2$ mixes non--perturbatively with an operator of lower dimension. 
Taking this effect into account leads to a considerably smaller value 
of $d_2$, which is consistent with the experimental data.
\end{abstract}
\pacs{12.38.Gc, 13.60.Hb, 13.88.+e}

\section{Introduction}

The nucleon's second spin--dependent structure function $g_2$ is of 
considerable phenomenological interest. The most important theoretical
tool for its analysis is the 
operator product expansion (OPE)~\cite{Jaffe}. In leading order 
of $1/Q^2$, $g_2$ receives contributions from both twist--two and twist--three
operators. It thus offers the unique possibility of directly assessing
higher--twist effects. The twist--three operator probes the
transverse momentum distribution of the quarks in the nucleon, and has 
no simple parton model interpretation.

In leading order of $1/Q^2$, and for massless quarks, the moments of $g_2$ 
are given by
\begin{equation}
\begin{split}
2\int_0^1\!\mbox{d}x \, x^n g_2(x,Q^2)  
  = \frac{1}{2}\frac{n}{n+1} \sum_{f=u,d} 
&\big[e^{(f)}_{2,n}(\mu^2/Q^2,g(\mu^2)) \: d_n^{(f)}(\mu) \\
&-e^{(f)}_{1,n}(\mu^2/Q^2,g(\mu^2))\: a_n^{(f)}(\mu)\big]
\end{split}
\label{mom}
\end{equation}
for even $n \ge 2$ in the flavor--nonsinglet sector. Here $f$ runs over the
light quark flavors. The reduced matrix
elements $a_n^{(f)}(\mu)$ and $d_n^{(f)}(\mu)$, taken in a nucleon state
with momentum $p$ and spin vector $s$, are defined by \cite{Jaffe}
\begin{eqnarray}
\langle p,s| 
 {\cO}^{5 (f)}_{ \{ \sigma\mu_1\cdots\mu_n \} }
                       | p,s \rangle 
   &=& \frac{1}{n+1}a_n^{(f)} \,[ s_\sigma p_{\mu_1} \cdots p_{\mu_n} 
+ \cdots -\mbox{traces}] \,, \label{twist2} \\
\langle p,s| 
 {\cO}^{5 (f)}_{ [  \sigma \{ \mu_1 ] \cdots \mu_n \} }
                       | p,s \rangle 
   &=& \frac{1}{n+1}d_n^{(f)} \,[ (s_\sigma p_{\mu_1} - s_{\mu_1} p_\sigma)
 p_{\mu_2}\cdots p_{\mu_n} + \cdots -\mbox{traces}] \,, 
\label{twist3} \\
 {\cO}^{5 (f)}_{\sigma\mu_1\cdots\mu_n}
   & =& \left(\frac{\mathrm i}{2}\right)^n\bar{\psi}\gamma_{\sigma} \gamma_5
  \Dd{\mu_1} \cdots \Dd{\mu_n}
 \psi -\mbox{traces} \,.
\end{eqnarray}
Here $\mu$ denotes the renormalization scale. 
The Wilson coefficients $e_{1,n}^{(f)}$, $e_{2,n}^{(f)}$ depend on the 
ratio of scales $\mu^2/Q^2$ and on the running coupling constant 
$g(\mu^2)$. The tree level values of the Wilson coefficients for
electroproduction are given by the quark charges $Q^{(f)}$:
\begin{equation} \label{wilcoeff}
e^{(f)}_{i,n} = Q^{(f)2}(1 + O(g^2)) \,. 
\end{equation}
The symbol $\{\cdots\}$ ($[\cdots]$) indicates symmetrization
(antisymmetrization) with 
\begin{equation} 
 {\cO}_{ \{ \mu_1\cdots\mu_n \} } = \frac{1}{n!} \sum_{\pi \in {\mathcal S}_n}
  {\cO}_{\mu_{\pi(1)} \cdots \mu_{\pi(n)}} \,.
\end{equation}
The operator (\ref{twist2}) has twist two, whereas
the operator (\ref{twist3}) has twist three.
The twist--two contribution in (\ref{mom}) is also known as the 
Wandzura--Wilczek contribution \cite{WW}.

Note for comparison that in leading order of $1/Q^2$ the moments of $g_1$ 
are given by the twist--two matrix elements $a_n^{(f)}$:
\begin{equation} \label{g1moments}
2\int_0^1\!\mbox{d}x \, x^n g_1(x,Q^2)  
  = \frac{1}{2} \sum_{f=u,d} e^{(f)}_{1,n}(\mu^2/Q^2,g(\mu^2))
\: a_n^{(f)}(\mu) \,. 
\end{equation}

Both the Wilson coefficients and the operators are renormalized at the
scale $\mu$. It is assumed that the Wilson coefficients can be computed 
perturbatively.
The reduced matrix elements $a_n^{(f)}$ and $d_n^{(f)}$, on the other
hand, are non--perturbative quantities and hence a problem for the lattice.
(Note that some authors use a different definition of $a_n$ 
and $d_n$, e.g.\ the values given in refs.~\cite{abe,anthony} have to 
be multiplied by 2 to agree with our conventions.)
In the following we shall drop the flavor indices, unless they are 
necessary.

A few years ago we computed the lowest non--trivial moment of $g_2$ on
the lattice~\cite{QCDSF1}. This calculation splits into two separate tasks.
The first task is to compute the nucleon matrix elements of the 
appropriate lattice operators. This was described in detail in~\cite{QCDSF1}.
The second task is to renormalize the operators. 
Renormalization effects are a major source of systematic error. An essential 
feature of our previous calculation was that the renormalization was done in 
perturbation theory and hence mixing with lower--dimensional operators
could not be taken into account. In that approach
the twist--three contribution turned out to be the
dominant contribution to both the proton and the neutron structure functions.
This result has been recently confirmed by Dolgov et al.\
\cite{dolgov}.

In the meantime, it has become possible to study renormalization 
non--perturbatively on the lattice, see e.g.~\cite{marti,jansen}.
This approach allows us to consider mixing with lower--dimensional operators.
If present, it will be the dominant mixing effect in the continuum limit.
Since the twist--three operators (\ref{twist3}) can suffer from such mixing,
we shall extend our previous work by employing non--perturbative 
renormalization. 
In a recent paper~\cite{QCDSF3} we have started a non--perturbative 
calculation of the renormalization constants associated with the 
structure functions $F_1$, $F_2$ and $g_1$ in the flavor--nonsinglet sector.  
Here we consider the case of the structure function $g_2$
restricting ourselves to $n = 2$, the 
lowest moment of $g_2$ for which the OPE makes a statement.  
A preliminary version of this work based on lower statistics at
a single value of the bare coupling has already been presented
in Ref.~\cite{g2rev}.

\section{Renormalization and Mixing in Continuum Perturbation Theory}

The renormalization of the operators which contribute to the moments of
$g_2$ has been studied by several authors in continuum perturbation 
theory \cite{contpert}. Since the more recent paper by Kodaira et al.\ 
\cite{Kodaira2} (see also \cite{Kodaira3}) is closest to the methods 
applied on the lattice, let us briefly recapitulate the main findings 
of these authors.

They consider the case $n=2$ in the flavor--nonsinglet sector
and start from the operators
\begin{eqnarray} 
 R_F^{\sigma \mu \nu} & = & \frac{\mbox{i}^2}{3} \left[ 
 2 \bar{\psi} \gamma_5 \gamma^\sigma D^{\{ \mu} D^{\nu \}} \psi
 - \bar{\psi} \gamma_5 \gamma^\mu D^{\{ \sigma} D^{\nu \}} \psi
 - \bar{\psi} \gamma_5 \gamma^\nu D^{\{ \mu} D^{\sigma \}} \psi \right]
 - \mbox{traces} \,,
\\ 
 R_1^{\sigma \mu \nu} & = & \frac{1}{12} g \left[ 
\epsilon^{\sigma \mu \alpha \beta}
 \bar{\psi} F_{\alpha \beta} \gamma^{\nu} \psi
+ \epsilon^{\sigma \nu \alpha \beta}
 \bar{\psi} F_{\alpha \beta} \gamma^{\mu} \psi \right] - \mbox{traces} \,,
\\ 
 R_m^{\sigma \mu \nu} & = & \mbox{i}  m
 \bar{\psi} \gamma_5 \gamma^\sigma D^{\{ \mu} \gamma^{\nu \}} \psi
 - \mbox{traces} \,,
\\ 
 R_{\mathrm {eq}}^{\sigma \mu \nu} & = & \frac{\mbox{i}}{3}  \left[ 
 \bar{\psi} \gamma_5 \gamma^\sigma D^{\{ \mu} \gamma^{\nu \}} 
  (\mbox{i} \DD - m) \psi
 +  \bar{\psi} (\mbox{i} \DD - m) 
    \gamma_5 \gamma^\sigma D^{\{ \mu} \gamma^{\nu \}} \psi \right]
 - \mbox{traces} \,.
\end{eqnarray}
Here $F_{\alpha \beta}$ denotes the gluon field strength tensor, which
could alternatively be expressed as a commutator of two covariant
derivatives. Due to the relation 
\begin{equation} 
 R_F^{\sigma \mu \nu} = \frac{2}{3} R_m^{\sigma \mu \nu} +  
 R_1^{\sigma \mu \nu} +  R_{\mathrm {eq}}^{\sigma \mu \nu} 
\end{equation}
it is possible to eliminate one of the above operators. A one--loop
calculation of the quark--quark--gluon three--point functions with a
single insertion of each of these operators reveals the necessity of taking 
one more operator into account in the process of renormalization, namely
the gauge--variant operator
\begin{equation}
 R_{\mathrm {eq1}}^{\sigma \mu \nu} = \frac{\mbox{i}}{3}  \left[ 
 \bar{\psi} \gamma_5 \gamma^\sigma \partial^{\{ \mu} \gamma^{\nu \}} 
  (\mbox{i} \DD - m) \psi
 +  \bar{\psi} (\mbox{i} \DD - m) 
    \gamma_5 \gamma^\sigma \partial^{\{ \mu} \gamma^{\nu \}} \psi \right]
 - \mbox{traces} \,.
\end{equation}
Of course, in physical matrix elements neither $R_{\mathrm {eq}}$ nor
$R_{\mathrm {eq1}}$ will contribute. They show up, however, in off--shell
vertex functions and influence the renormalization factors.

Kodaira et al.\ choose $R_1$ and $R_m$ as the physical operators. In the 
chiral limit $m \to 0$ $R_m$ is neglected, and they obtain for the 
scale dependence of the twist--three piece
\begin{equation} \label{d2rg}
\int_0^1 \! {\mathrm d}x \, x^2 g_2^{\mathrm {twist-3}}(x,Q^2) =
\left( \frac{\alpha_s (Q^2)}{\alpha_s (\mu^2)} \right)^\omega
\int_0^1 \! {\mathrm d}x \, x^2 g_2^{\mathrm {twist-3}}(x,\mu^2) \,,
\end{equation}
where for $N_c$ colors and $N_f$ flavors
\begin{equation} 
\omega = \frac{3 N_c - \frac{1}{3}\frac{N_c^2 - 1}{2 N_c}}
              {\frac{11}{3}N_c - \frac{2}{3}N_f } 
\end{equation}
in agreement with earlier calculations.

Using $R_F$ and $R_1$ as the physical operators, one finds that in the 
large--$N_c$ limit the operator $R_F$ dominates the renormalization 
group evolution of the nucleon matrix elements. 
This has been shown by Ali, Braun, and Hiller~\cite{ali},
and was rederived in the present framework by Sasaki~\cite{Sasaki}.

\section{Renormalization and Mixing on the Lattice}

In a lattice calculation, the first step is the analytic continuation
to imaginary times, leading from the physical Minkowski space to
Euclidean space. From now on, all expressions are written for the
Euclidean case (for the details of our conventions see Appendix A of
Ref.~\cite{cbest}). Hence we have to study operators of the form
\begin{equation}
  {\cO} ^5_{\sigma \mu_1 \cdots \mu_n} = 2^{-n}  
     \bar{\psi}\gamma_\sigma \gamma_5 \Dd{\mu_1} \cdots \Dd{\mu_n} \psi \,.
\end{equation}
We shall neglect quark masses, i.e.\ we consider only the chiral limit.
In our earlier work~\cite{QCDSF1,QCDSF2} we have computed the renormalization
constants in perturbation theory to one--loop order. However, it is 
believed that perturbation theory cannot give reliable values for the 
mixing with lower--dimensional operators because non--perturbative 
effects are expected to be important.

For a multiplicatively renormalizable operator, i.e.\ in the absence
of mixing, we can write
\begin{equation}
   {\cO}_{\mathrm R}(\mu) = Z_{\cO}(a\mu)\, {\cO}(a) \,,
\label{op1}
\end{equation}
where $a$ is the lattice spacing.
The renormalization constant $Z_{\cO}$ is fixed by a suitable condition.
As in the continuum, we impose the (MOM--like) renormalization condition
\begin{equation}  \label{MOM1}
 {\rm tr} \left( \Gamma_{\mathrm R} (p)
   \Gamma_{\mathrm {Born}} (p) ^\dagger \right) \underset{p^2 =\mu^2}{=}
 {\rm tr} \left( \Gamma_{\mathrm {Born}} (p)
   \Gamma_{\mathrm {Born}} (p) ^\dagger \right) 
\end{equation}
on the corresponding quark--quark vertex function in the Landau gauge. 
Here $\Gamma_{\mathrm {Born}} (p)$ denotes the Born or tree--level
contribution to the vertex function. 
The renormalized vertex function $\Gamma_{\mathrm R} (p)$ and its
bare precursor $\Gamma (p)$ are related by multiplicative renormalization:
\begin{equation} 
\Gamma_{\mathrm R} (p) = Z_\psi^{-1} Z_{\cO} \Gamma (p) \,,
\end{equation}
where $Z_\psi = Z_\psi (a \mu)$ is the quark wave function renormalization
constant defined as in Ref.~\cite{QCDSF3}.

As before~\cite{QCDSF1}, we give the nucleon a momentum in the 1--direction
and choose the polarization in the 2--direction. With these choices we use
the operator
\begin{equation}
{\cO}^5_{\{214\}} =: {\cO}^{\{5\}}
\label{oa2}
\end{equation}
for the twist--two matrix element $a_2$. It belongs to the 
representation $\tau_3^{(4)}$ of the hypercubic group 
$H(4)$~\cite{Mandula,group}
and this property protects it from mixing with operators of equal
or lower dimension. Hence it is multiplicatively renormalizable, and the
operator renormalized at the scale $\mu$ is written as
\begin{equation}
{\cO}^{\{5\}}_{\mathrm R}(\mu) = 
      Z^{\{5\}}(a\mu) {\cO}^{\{5\}}(a) \,.
\end{equation}

As the operator for the twist--three
matrix element $d_2$ we take
\begin{eqnarray}
{\cO}^5_{[2\{1] 4\}} &=&  \third \left( 2 {\cO}^5_{2\{14\}} 
- {\cO}^5_{1\{24\}} - {\cO}^5_{4\{12\}} \right) \nonumber \\   
&=& \mbox{\small $\frac{1}{12}$}
   \bar{\psi}\Big(\gamma_2 \Dd{1} \Dd{4} + \gamma_2 \Dd{4} \Dd{1} 
   - \half \gamma_1 \Dd{2} \Dd{4} - \half \gamma_1 \Dd{4} \Dd{2} \nonumber \\
& & \quad \quad \quad - \half \gamma_4 \Dd{1} \Dd{2} - \half \gamma_4 \Dd{2}
\Dd{1}\Big) \gamma_5 \psi \nonumber \\
&=:& {\cO}^{[5]} \,, \label{o5}
\end{eqnarray}
which belongs to the representation $\tau_1^{(8)}$ of $H(4)$.
The operator (\ref{o5}) has dimension five and
$C$--parity $+$ and is the Euclidean counterpart of the Minkowski operator
$R_F$. It turns out that there exist two more operators of dimension
four and five, respectively, transforming identically 
under $H(4)$ and having the same $C$--parity, with which (\ref{o5}) can mix:
\begin{equation}
\mbox{\small $\frac{1}{12}$} \mbox{i}\, 
 \bar{\psi} \Big(\sigma_{13} \Dd{1} -  \sigma_{43} \Dd{4}\Big) \psi
=: {\cO}^\sigma, \label{osigma}
\end{equation}
\vspace*{-0.7cm}
\begin{equation}
\mbox{\small $\frac{1}{12}$}
\bar{\psi} \Big(\gamma_1 \Dd{3} \Dd{1} - \gamma_1 \Dd{1}
 \Dd{3} - \gamma_4 \Dd{3} \Dd{4} + \gamma_4 \Dd{4} \Dd{3}\Big) \psi =:
{\cO}^0. \label{o0}
\end{equation}
We use the definition 
$\sigma_{\mu \nu} = (\mathrm i/2) [\gamma_\mu , \gamma_\nu ]$.

The operator ${\cO}^0$ is the Euclidean analog of $R_1$ with the field
strength replaced by a commutator of two covariant derivatives, and 
${\cO}^\sigma$ corresponds to $R_m$. In continuum perturbation
theory $R_m$ can be neglected in the chiral limit. On the lattice, 
the explicit breaking of chiral symmetry induced by Wilson--type 
fermions, which we shall use, persists
even when the quarks are massless. For dimensional reasons,
we expect that ${\cO}^\sigma$ contributes with a coefficient 
$\propto a^{-1}$ and hence has to be kept. The operator ${\cO}^0$, 
on the other hand, being of the same dimension as ${\cO}^{[5]}$, mixes 
with a coefficient of order $g^2$, which should be small. Therefore
we discard ${\cO}^0$ as well as possible lattice counterparts of 
$R_{\mathrm {eq}}$ and $R_{\mathrm {eq1}}$, which are also of dimension 
five and hence are also multiplied by a factor of order $g^2$. 
The above mentioned observation \cite{ali,Sasaki} that $R_F$ 
dominates over $R_1$ in the renormalization group evolution as 
$N_c \to \infty$ may be taken as another indication that neglecting 
${\cO}^0$ is not unreasonable. However, this dominance holds only in
physical matrix elements and does not apply to the mixing with the operators
$R_{\mathrm {eq}}$ and $R_{\mathrm {eq1}}$.

So we make the following ansatz for ${\cO}^{[5]}$ renormalized 
at the scale $\mu$: 
\begin{equation}
{\cO}^{[5]}_{\mathrm R}(\mu) = Z^{[5]}(a\mu) {\cO}^{[5]}(a) + 
  \frac{1}{a} Z^\sigma(a\mu) {\cO}^\sigma(a) \,.
\label{renorm}
\end{equation}
The renormalization constant $Z^{[5]}$ and the mixing coefficient $Z^\sigma$
are determined from the conditions 
\begin{eqnarray}
  {\rm tr} \left( \Gamma^{[5]}_{\mathrm R} (p)
   \Gamma^{[5]}_{\mathrm {Born}} (p) ^\dagger \right) 
                                 & \underset{p^2 =\mu^2}{=} &
  {\rm tr} \left( \Gamma^{[5]}_{\mathrm {Born}} (p)
   \Gamma^{[5]}_{\mathrm {Born}} (p) ^\dagger \right)  \,,
\\
  {\rm tr} \left( \Gamma^{[5]}_{\mathrm R} (p)
   \Gamma^\sigma_{\mathrm {Born}} (p) ^\dagger \right) 
                                 & \underset{p^2 =\mu^2}{=} &
  {\rm tr} \left( \Gamma^{[5]}_{\mathrm {Born}} (p)
   \Gamma^\sigma_{\mathrm {Born}} (p) ^\dagger \right) = 0 \,,
\end{eqnarray}
which are straightforward generalizations of Eq.~(\ref{MOM1}).
Note that the operator (\ref{o0}) vanishes in the Born approximation 
between quark states, which is another reason why we do not take it
into account.

Rewriting Eq.~(\ref{renorm}) as
\begin{equation}
{\cO}^{[5]}_{\mathrm R} (\mu) = Z^{[5]}(a\mu) \left( {\cO}^{[5]}(a) + 
\frac{1}{a} \frac{Z^\sigma(a\mu)}{Z^{[5]}(a\mu)} {\cO}^\sigma(a) \right) 
\end{equation}
we see that ${\cO}^{[5]}(\mu)$ will have a multiplicative dependence 
on $\mu$ (cf.\ Eq.~(\ref{op1}))
only if the ratio $Z^\sigma(a\mu)/Z^{[5]}(a\mu)$ does not
depend on $\mu$. The scale dependence will then completely reside
in $Z^{[5]}$.

\section{Simulation details}

We have obtained numerical results for matrix elements and $Z$ factors
in quenched simulations at $\beta = 6/g_0^2 = 6.0$, 6.2, and 6.4
($g_0 =$ bare coupling constant on the lattice).
Whereas our original calculation~\cite{QCDSF1} 
at $\beta = 6.0$ used Wilson fermions, we have meanwhile 
switched to non--perturbatively improved fermions (clover fermions)
in order to reduce $O(a)$ effects. The value of the clover coefficient 
$c_{\mathrm{SW}}$ is taken from Ref.~\cite{alpha}.
Since we have not improved the
operators, there will still be residual $O(a)$ effects in the matrix
elements and the renormalization factors. A few details
of our computations are collected in Table~\ref{param}.

The matrix elements are calculated from several hundred configurations
for each $\beta$. To compute the renormalization factors we use 
a momentum source~\cite{QCDSF3}. Therefore the statistical error is
$\propto (V N_{\mathrm {conf}})^{-1/2}$ for $N_{\mathrm {conf}}$ 
configurations on a lattice of volume $V$, and we already get small 
statistical uncertainties even from a small number of 
configurations, four in our case. (There is of course a price to be paid, 
the calculation for each momentum is independent, so the 
number of inversions of the fermion matrix is proportional to the number 
of momentum values.) The main source of statistical uncertainty in our
final results is from the matrix elements, not the $Z$ values.

The momenta in the vertex functions used for the evaluation of the
renormalization factors have been chosen close to the diagonal in the
Brillouin zone in order to keep cut--off effects as small as possible.
One should bear in mind that this diagonal extends up to 
$p^2 = 4 \pi^2 / a^2$, but we use only momenta with $p^2 < \pi^2 / a^2$.

In each case, the calculations are done at three (or more) values of 
the hopping parameter $\kappa$ determining the bare quark mass
so that we can extrapolate our results (both the bare matrix elements and
renormalization factors) to the chiral limit. The extrapolation is
performed linearly in $m_\pi^2$, the square of the pion mass.

The bare reduced matrix elements are calculated from three--point functions
in the standard fashion (see, e.g., Ref.~\cite{QCDSF1}). 
In Eq.(25) of Ref.~\cite{QCDSF1} the ratios of three-- to two--point
functions for the $a_2$ operator ${\cO}^{\{5\}}$ and the $d_2$ operator 
${\cO}^{[5]}$ are given. For the operators ${\cO}^0$ and ${\cO}^\sigma$
the ratios and ratio factors are the same as for the $d_2$ operator.
The matrix elements are collated
in Tables~\ref{rmeclo60}, \ref{rmeclo62}, \ref{rmeclo64} 
separately for $u$ and $d$ quarks in the proton. Here $d_2^{[5]}$ 
and $d_2^{\sigma}$ 
correspond to the operators (\ref{o5}) and (\ref{osigma}), respectively.
In addition we give the pion masses (in lattice units) which we use in the 
chiral extrapolations. They are mostly taken from Ref.~\cite{quarkmass}.
Note that all our errors are purely statistical. They were determined
by the jackknife procedure.

\section{Numerical Results for Renormalization Coefficients}

\begin{figure}
  \begin{center}
    \epsfig{file=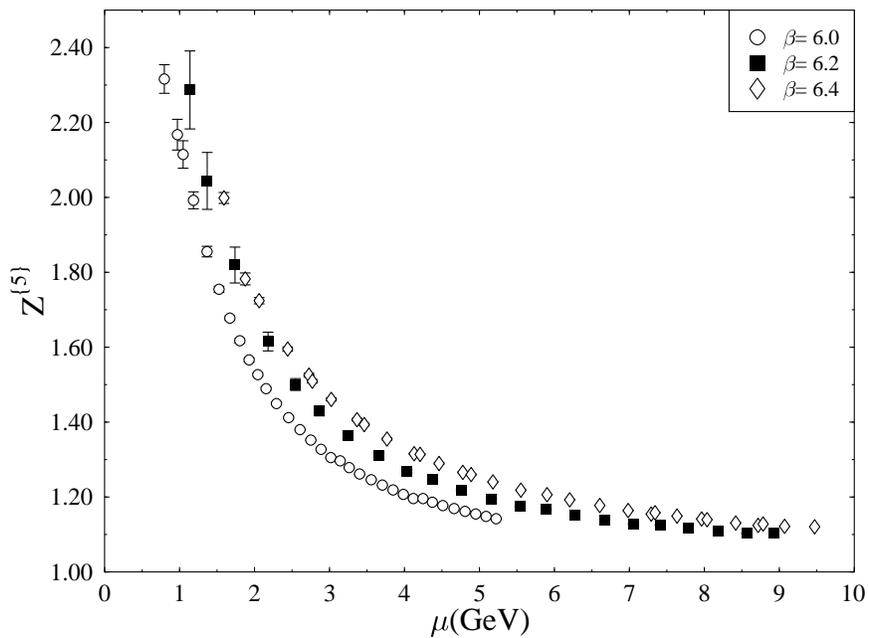,width=15cm}
    \caption{The renormalization constant $Z^{\{5\}}$ in the
             $\overline{\rm MS}$ scheme.}
    \label{fig.za2}
  \end{center}
\end{figure}

Let us begin the more detailed presentation of our numerical results with the 
renormalization factor $Z^{\{5\}}(a\mu)$ of the
multiplicatively renormalizable $a_2$ operator (\ref{oa2}). 
We convert our MOM numbers to the $\overline{\rm MS}$ scheme using
1--loop continuum perturbation theory as described in \cite{QCDSF3}.
In Fig.~\ref{fig.za2} we show the $\mu$ dependence of $Z^{\{5\}}$ 
extrapolated to the chiral limit. 
Results for Wilson fermions can be found in Ref.~\cite{QCDSF3}.
Note that at scales $\mu^2$ exceeding a few times the lattice 
cut--off $a^{-2}$ strong lattice artifacts may be present so that 
the corresponding results should not be taken too seriously.

\begin{figure}
 \vspace*{-2.5cm}
  \begin{center}
    \epsfig{file=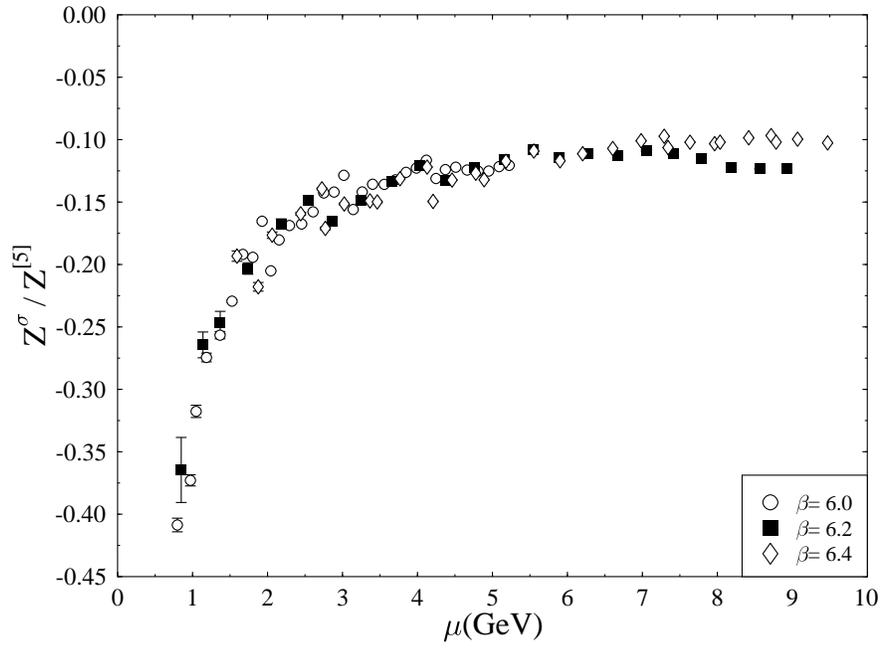,width=15cm}
    \caption{The ratio $Z^\sigma(a\mu)/Z^{[5]}(a\mu)$.}
    \label{fig.zrat}
  \end{center}
\end{figure}

Turning to the more subtle renormalization of the $d_2$ operator 
(\ref{o5}) we must note that
the conversion factor from our MOM scheme to the $\overline{\rm MS}$ scheme
has not yet been calculated because of the complications caused by
the mixing effects. 
Therefore we stick to the MOM numbers.
Let us first consider the ratio $Z^\sigma(a\mu)/Z^{[5]}(a\mu)$.
As discussed above, $Z^\sigma(a\mu)/Z^{[5]}(a\mu)$ should be 
independent of the renormalization 
scale $\mu$ if the renormalized operator is to depend on $\mu$ 
multiplicatively. In Fig.~\ref{fig.zrat} we show this ratio for 
our three $\beta$ values. It becomes approximately flat for
scales $\mu$ larger than about 3.5 GeV. 
While a scale of 3.5 GeV might seem to be 
somewhat too close to the cut--off for $\beta = 6.0$ and perhaps also
for $\beta = 6.2$, it enters the region where lattice artifacts die out  
in the case of $\beta = 6.4$. Therefore we feel encouraged to apply
the $Z$ factors around $\mu = 3.5 \, \mbox{GeV}$ in order to evaluate
structure function moments in the next section.

\begin{figure}
 \vspace*{-2.5cm}
  \begin{center}
    \epsfig{file=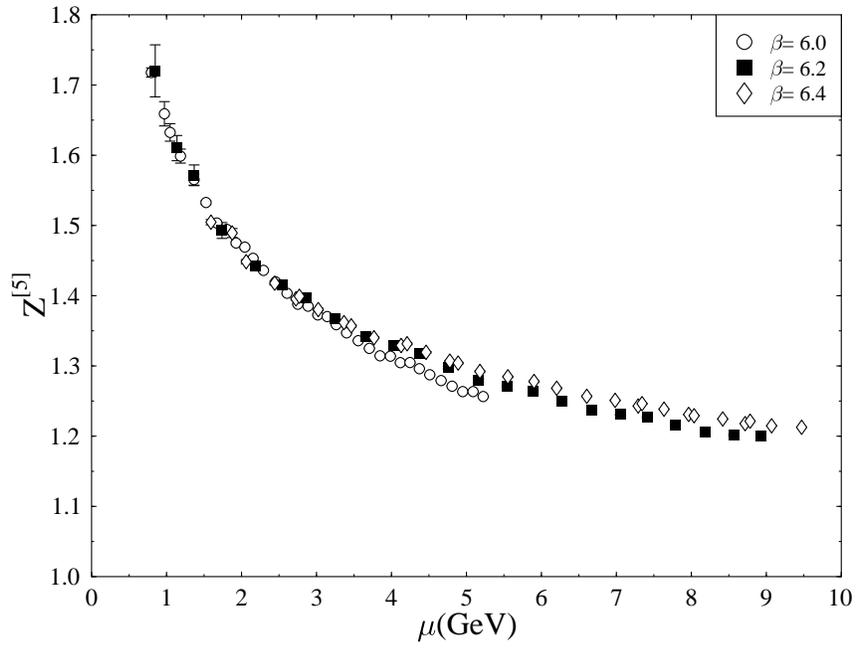,width=15cm}
    \caption{The renormalization factor $Z^{[5]}(a\mu)$.}
    \label{fig.z5}
  \end{center}
\end{figure}

In Fig.~\ref{fig.z5} we plot $Z^{[5]}(a\mu)$ in order to show the size 
of the multiplicative renormalization in our approach.

\section{Numerical Results for Structure Function Moments}

\begin{figure}
 \vspace*{-2.5cm}
  \begin{center}
    \epsfig{file=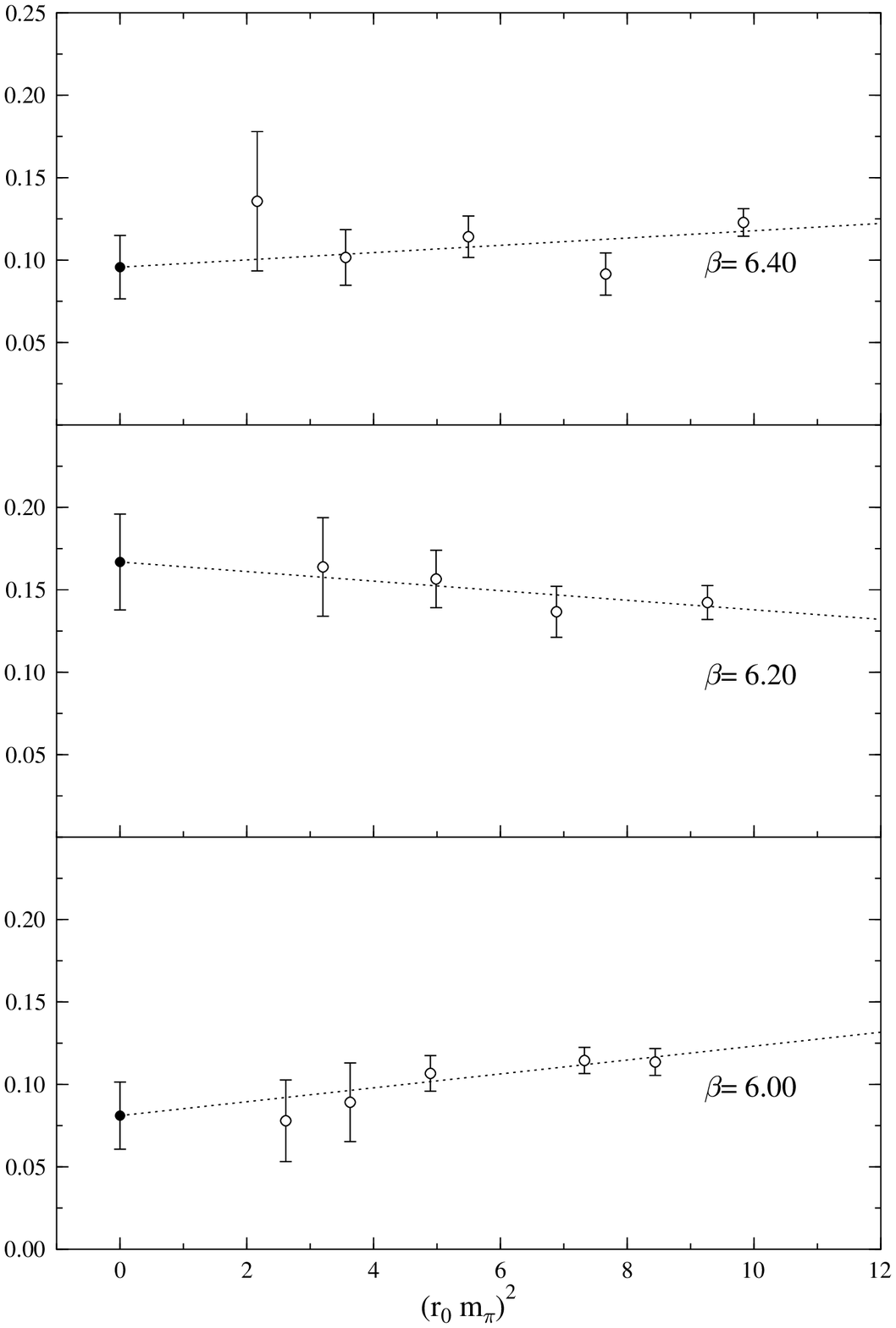,width=15cm}
    \caption{Chiral extrapolation of the bare matrix element $a_2^{(u)}$
             in the proton.}
    \label{fig.chiex.a2u}
  \end{center}
\end{figure}

\begin{figure}
 \vspace*{-2.5cm}
  \begin{center}
    \epsfig{file=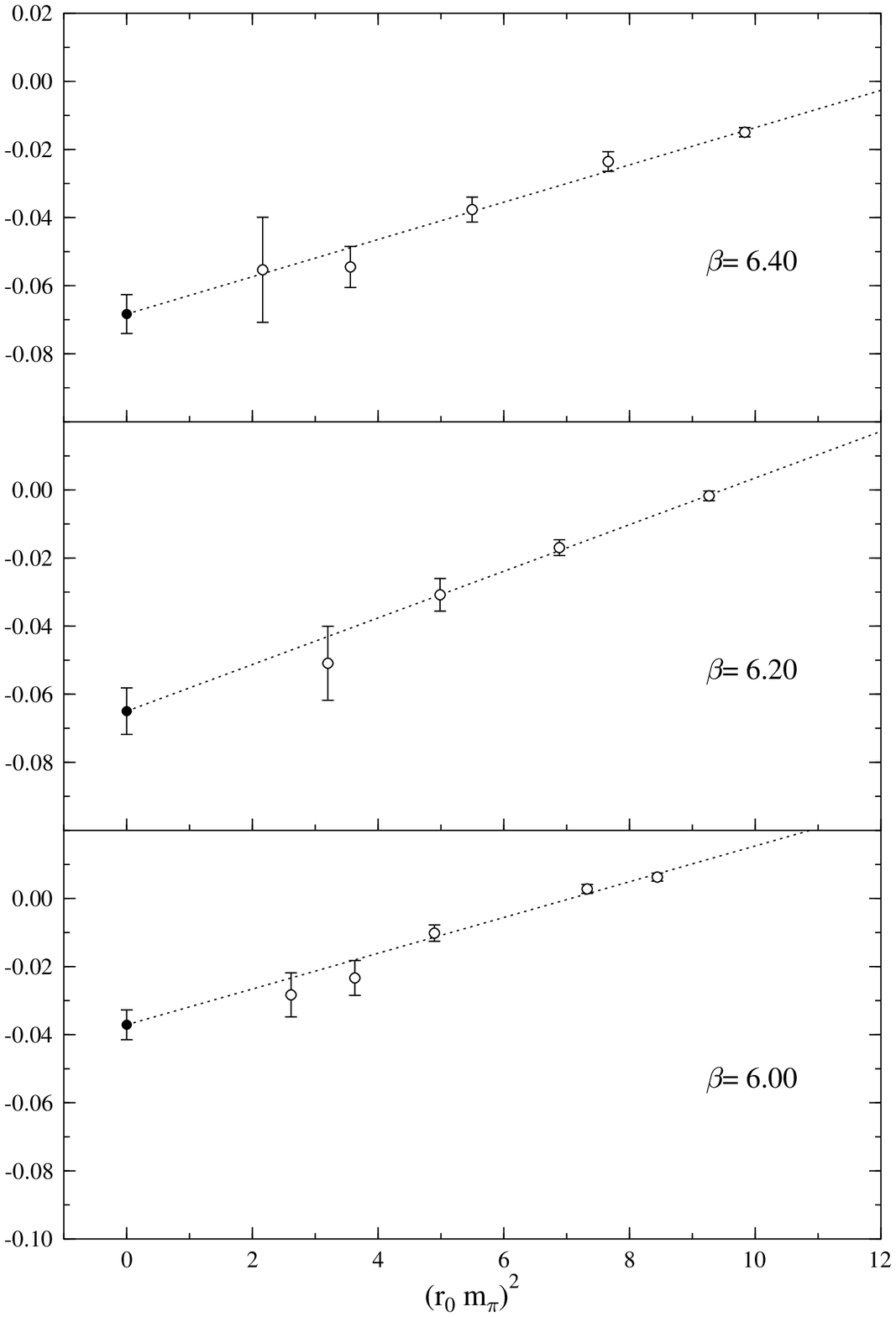,width=15cm}
    \caption{Chiral extrapolation of the bare matrix element 
             $ d_2^{[5]\,(u)} $ in the proton.}
    \label{fig.chiex.d25u}
  \end{center}
\end{figure}

Let us now discuss our nucleon matrix elements. 
In Figs.~\ref{fig.chiex.a2u}, \ref{fig.chiex.d25u} we show 
the chiral extrapolations of the bare values of $a_2^{(u)}$ and 
$ d_2^{[5]\,(u)} $, respectively. Unfortunately, like in all other
current QCD simulations our quark masses are
rather large so that the 
extrapolation has to bridge quite some gap. 
A striking feature of the data is that the bare $a_2^{(u)}$
values at $\beta = 6.2$ are rather different from the values at
the other two $\beta$'s. We can 
interpret this only as an unpleasantly large statistical fluctuation.
The bare matrix elements in the chiral limit will be combined with
the renormalization factors of the preceding section to yield 
estimates of the renormalized matrix elements. 

\begin{figure}
  \begin{center}
    \epsfig{file=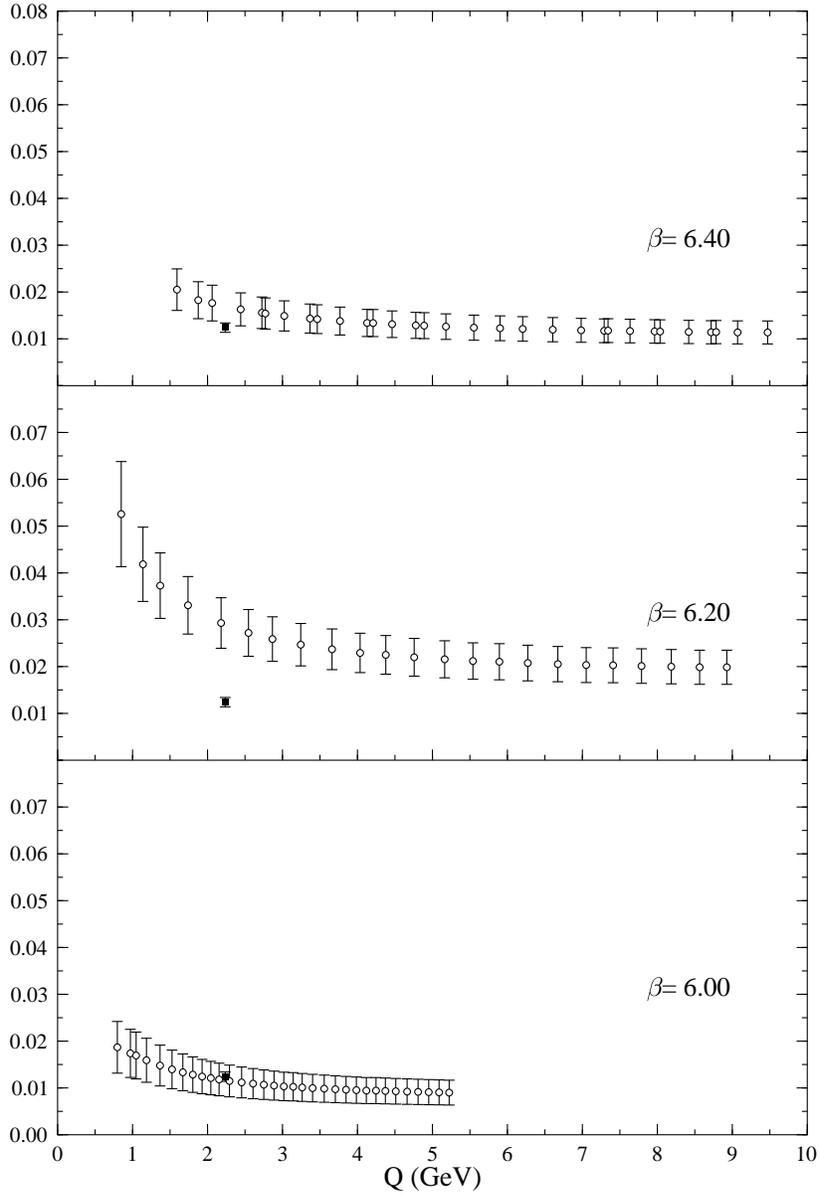,width=15cm}
    \caption[dummy]{The moment $\int_0^1\mbox{d}x x^2 g_1(x,Q^2)$ 
             for the proton.
             The square indicates the experimental value \cite{abe}.}
    \label{fig.g1}
  \end{center}
\end{figure}

We start with the twist--two matrix element $a_2$. In the 
$\overline{\rm MS}$ scheme with anticommuting $\gamma_5$ 
the corresponding Wilson coefficient is given by (see, e.g., Ref.~\cite{marco})
\begin{equation} \begin{array}{l} \displaystyle
 e^{(f)}_{1,2}(\mu^2/Q^2,g(\mu^2)) = Q^{(f)2} 
  \left( \frac{g^2(Q^2)}{g^2(\mu^2)} \right)^{\gamma_0/(2 \beta_0)}
\\  \displaystyle \qquad {}
 \times \left[ 1 + \frac{1}{16 \pi^2} \left(  g^2(Q^2) - g^2(\mu^2) \right)
 \left( \frac{\gamma_1}{2 \beta_0} - \frac{\gamma_0 \beta_1}{2 \beta_0^2}
   \right)  + \frac{g^2(Q^2)}{16 \pi^2} \frac{5}{3} \right] 
\end{array}
\end{equation}
with $\beta_0 = 11$, $\beta_1 = 102$, $\gamma_0 = 100/9$, 
$\gamma_1 = 141.78$. (These are the numbers for $N_f = 0$ flavors, appropriate
for the quenched approximation.) The renormalized reduced matrix element
is obtained from the bare value (extrapolated to the chiral limit) after
multiplication with the non--perturbative renormalization factor converted
to the $\overline{\rm MS}$ scheme. From Eq.~(\ref{g1moments}) we can then
calculate $\int_0^1\mbox{d}x x^2 g_1(x,Q^2)$. To avoid large logarithms
in the Wilson coefficient we put $Q^2 = \mu^2$. In Fig.~\ref{fig.g1} 
we show the results for the proton and compare with the experimental 
value \cite{abe}. While our lattice results at $\beta = 6.0$ agree 
surprisingly well with the experimental number, the above mentioned 
fluctuation makes them considerably larger at $\beta = 6.2$. Fortunately,
they drop again at $\beta = 6.4$. For the neutron there is a similar
effect, but due to the larger errors it is less significant.

Let us now turn to our results for the twist--3 matrix elements.
In this case it is unclear how to convert our MOM results
to the $\overline{\rm MS}$ scheme due to the mixing effects. Therefore
we do not make use of the $\overline{\rm MS}$ Wilson coefficient,
which has recently been calculated \cite{ji} (with the 
't Hooft--Veltman $\gamma_5$). It would change the final results 
for $d_2$ by $\approx 10 \%$. Instead we use only
the lowest--order approximation for the coefficient functions, 
i.e.\ the tree--level coefficients (\ref{wilcoeff}), which are the same
in all schemes, and define (by a slight abuse of notation)
\begin{eqnarray}
 d_2^{(p)} & = & Q^{(u)2} d_2^{(u)} + Q^{(d)2} d_2^{(d)} \,, \\
 d_2^{(n)} & = & Q^{(d)2} d_2^{(u)} + Q^{(u)2} d_2^{(d)} 
\end{eqnarray}
for the proton and the neutron, respectively. The renormalized values of
$d_2^{(f)}$ for $f=u,d$ in the proton are calculated from
\begin{equation}
 d_2^{(f)} = Z^{[5]} d_2^{[5]\,(f)} +
  \frac{1}{a} Z^\sigma d_2^{\sigma \,(f)} \,.
\end{equation}
Remember that, besides the twist--three matrix element $d_2$,
$\int_0^1\mbox{d}x x^2 g_2(x,Q^2)$ also contains a twist--two piece,
the Wandzura--Wilczek contribution, see Eq.~(\ref{mom}). To be consistent
we restrict ourselves to the tree--level Wilson coefficients and the 
MOM matrix elements also in this contribution when computing 
$\int_0^1\mbox{d}x x^2 g_2(x,Q^2)$ from (cf.\ Eqs.~(\ref{mom}) 
and (\ref{g1moments}))
\begin{equation}
\int_0^1\!\mbox{d}x \, x^2 g_2(x,Q^2) = \frac{1}{6} d_2 - 
  \frac{2}{3} \int_0^1\!\mbox{d}x \, x^2 g_1(x,Q^2) \,.  
\label{x2g2}
\end{equation}
 
\begin{figure}
  \begin{center}
    \epsfig{file=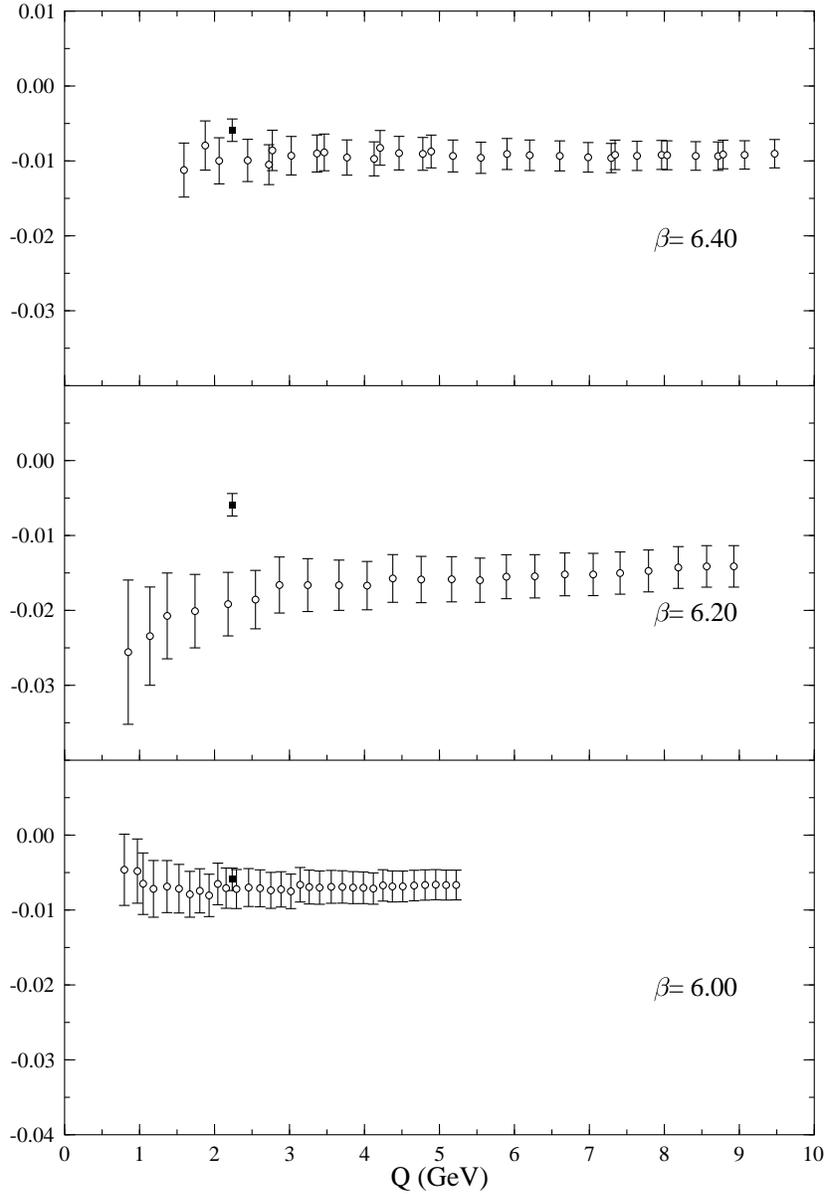,width=15cm}
    \caption[dummy]{The moment $\int_0^1\mbox{d}x x^2 g_2(x,Q^2)$ 
             for the proton.
             The square indicates the experimental value obtained by 
             combining results from \cite{abe} and \cite{anthony}.}
    \label{fig.g2}
  \end{center}
\end{figure}

The moment $\int_0^1\mbox{d}x x^2 g_2(x,Q^2)$ is plotted in Fig.~\ref{fig.g2}
for the proton, where we have again identified $Q^2 = \mu^2$.
The experimental value is obtained by combining 
$\int_0^1\mbox{d}x x^2 g_1(x,Q^2)$ from Ref.~\cite{abe} with $d_2$
from Ref.~\cite{anthony}. Again we see the effect of the ``fluctuation''
at $\beta = 6.2$.

\begin{figure}
  \begin{center}
    \epsfig{file=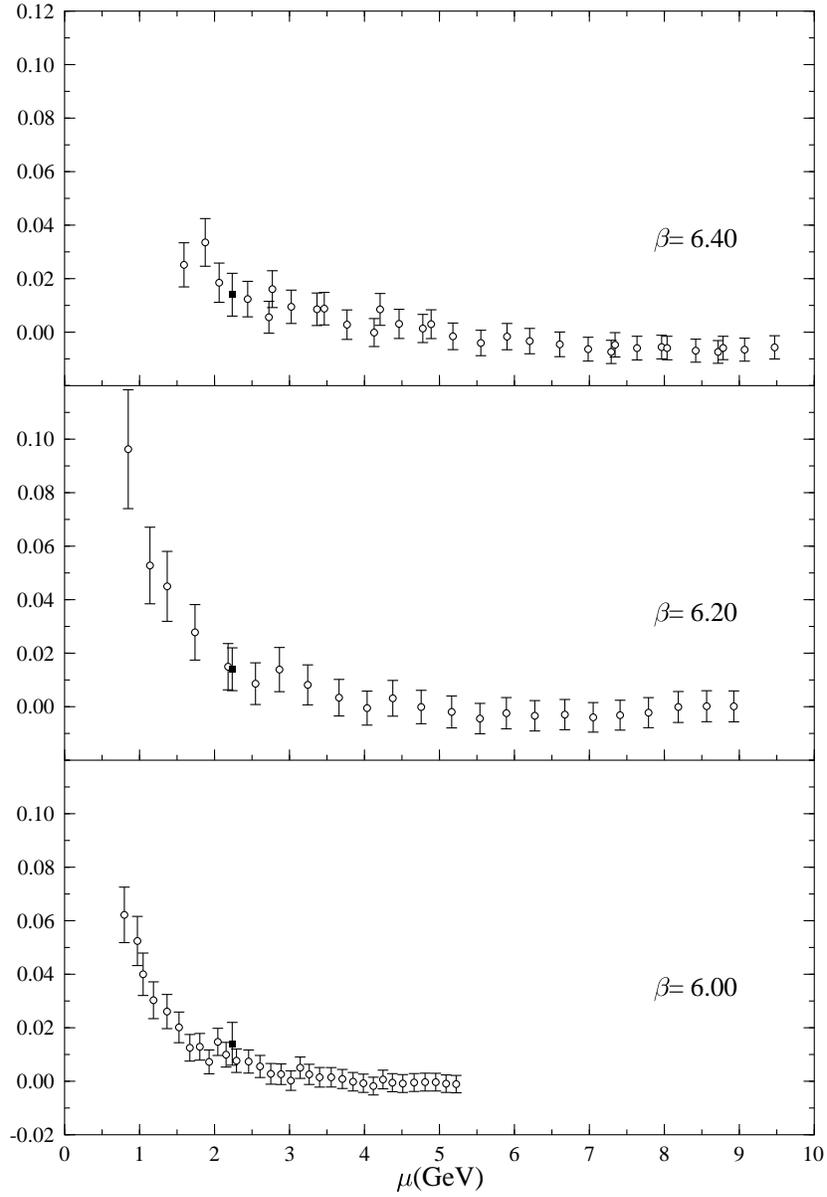,width=15cm}
    \caption[dummy]{The reduced matrix element $d_2$ in the proton.
             The square indicates the experimental value~\cite{anthony}.}
    \label{fig.d2}
  \end{center}
\end{figure}

Comparing the proton results shown in Fig.~\ref{fig.g2} with 
the numbers presented in Fig.~\ref{fig.g1} one sees that 
$\int_0^1\mbox{d}x x^2 g_2(x,Q^2)$ is dominated by the twist--two operator.
There is little room left for the twist--three operator, and
one obtains rather small values for $d_2$ as shown in Fig.~\ref{fig.d2}
for the proton. In the neutron, $d_2$ is even smaller in magnitude
and hardly different from zero within the statistical errors.

In Tables~\ref{res5} and \ref{res10} we present results at the scales 
5 GeV$^2$ and 10 GeV$^2$, respectively. Note that here 
$\int_0^1\mbox{d}x x^2 g_1(x,Q^2)$ includes the one--loop Wilson
coefficient as well as the conversion factor to the $\overline{\rm MS}$ 
scheme, both of which were neglected in the calculation of 
$\int_0^1\mbox{d}x x^2 g_2(x,Q^2)$ for the reason explained above.
The difference between the two sides of Eq.~(\ref{x2g2}) when evaluated
with the numbers taken from the tables gives therefore an impression
of the uncertainties originating from our incomplete knowledge of the
perturbative corrections.

\begin{figure}
  \begin{center}
    \epsfig{file=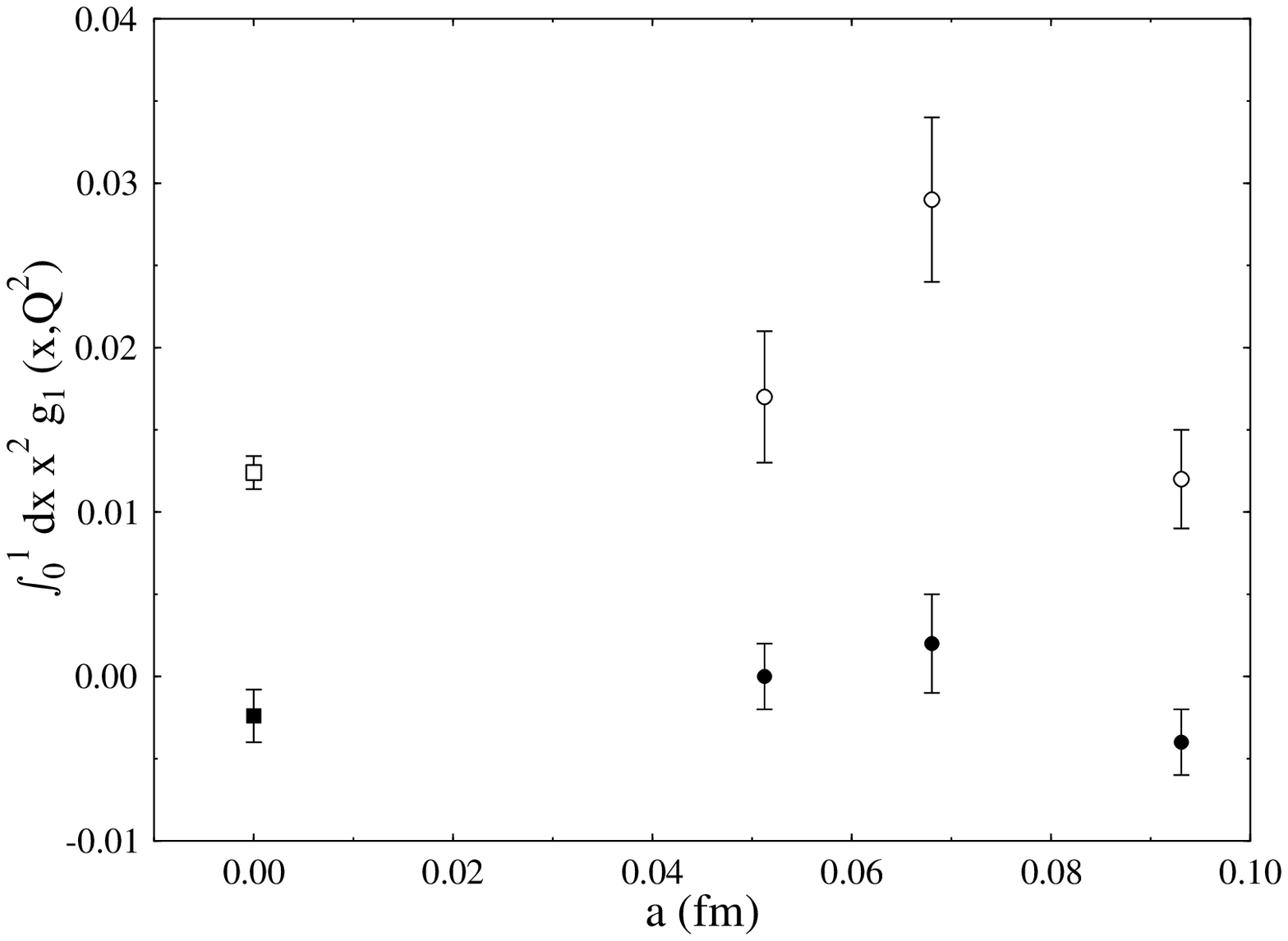,width=15cm}
    \caption[dummy]{The moment $\int_0^1\mbox{d}x x^2 g_1(x,Q^2)$ at 
             $Q^2 = 5 \, \mbox{GeV}^2$ for the proton (open symbols) 
	     and the neutron (filled symbols) plotted versus the lattice
	     spacing $a$. The squares at $a=0$ indicate the experimental
	     values \cite{abe}.}
    \label{fig.contlim.g1}
  \end{center}
\end{figure}

\begin{figure}
  \begin{center}
    \epsfig{file=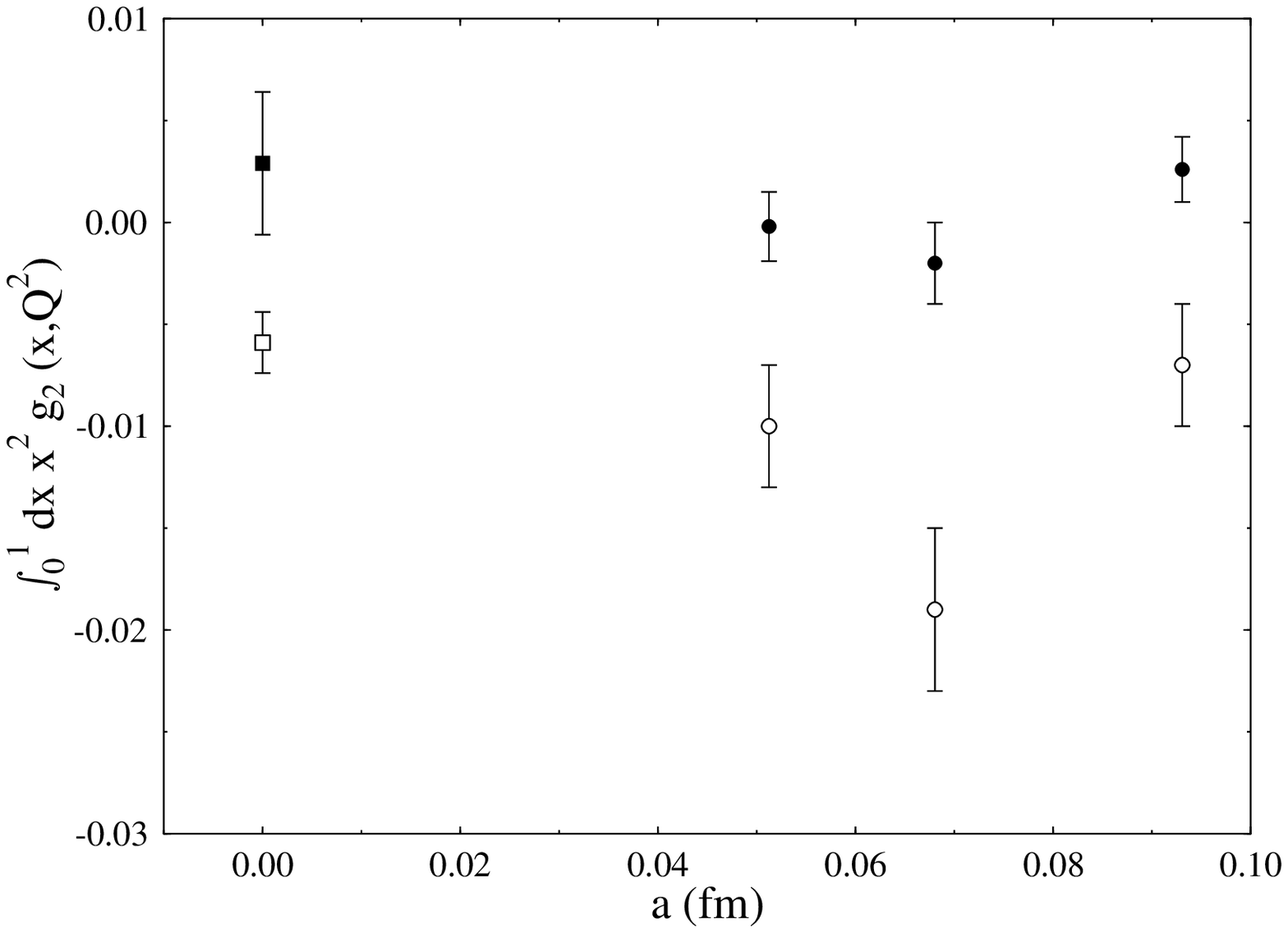,width=15cm}
    \caption[dummy]{The moment $\int_0^1\mbox{d}x x^2 g_2(x,Q^2)$ at 
             $Q^2 = 5 \, \mbox{GeV}^2$ for the proton (open symbols) 
	     and the neutron (filled symbols) plotted versus the lattice
	     spacing $a$. The squares at $a=0$ indicate the experimental
	     values obtained by combining results from  \cite{abe} and 
             \cite{anthony}.}
    \label{fig.contlim.g2}
  \end{center}
\end{figure}

\begin{figure}
  \begin{center}
    \epsfig{file=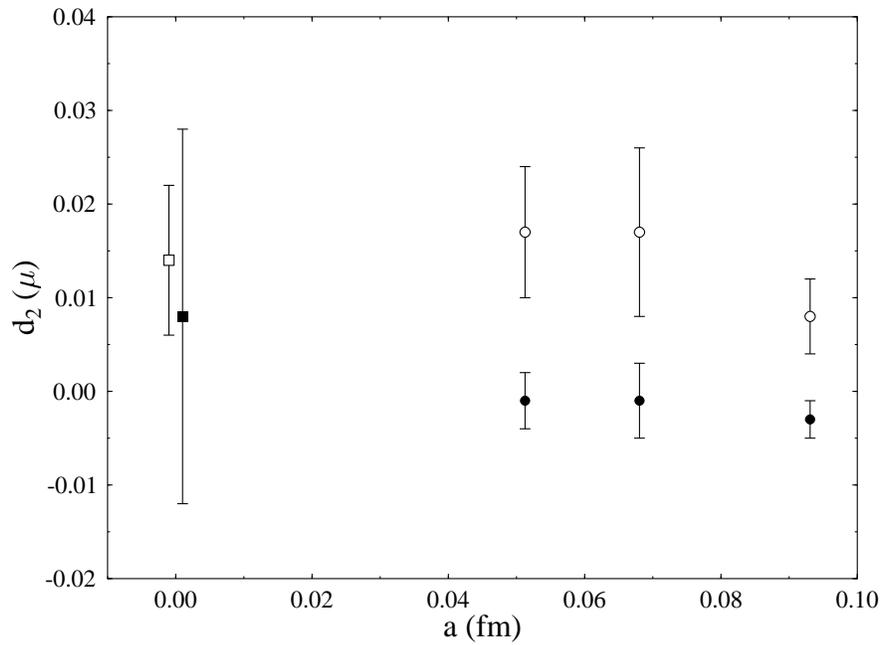,width=15cm}
    \caption[dummy]{The reduced matrix element $d_2$ at 
             $\mu^2 = 5 \, \mbox{GeV}^2$ for the proton (open symbols) 
	     and the neutron (filled symbols) plotted versus the lattice
	     spacing $a$. The squares at $a=0$ indicate the experimental
	     values~\cite{anthony}. They are plotted with a slight 
             horizontal offset to avoid overlapping error bars.}
    \label{fig.contlim.d2}
  \end{center}
\end{figure}

In Figs.~\ref{fig.contlim.g1}, \ref{fig.contlim.g2}, \ref{fig.contlim.d2}
we fix the scale at 5 GeV$^2$ and plot our results for the 
proton as well as for the neutron versus the lattice spacing $a$. 
Although an extrapolation to the continuum limit appears to be 
problematic, it is reassuring to see that we are getting close to
the experimental numbers shown at $a=0$. 

Of course, we should not forget that our computation suffers from various 
uncertainties.  
Apart from the fact that our treatment of the operator mixing is still
incomplete, these concern, e.g., the influence of the 
quenched approximation, the extrapolation to the chiral limit, and the size of
the lattice artifacts.
Sea--quark effects are expected to be concentrated at small $x$, hence they
should be suppressed by the factor $x^2$ in the moment which we have 
considered. Therefore we may hope that the quenched
approximation is reasonable in the case at hand. If indeed the valence quarks
dominate, then it should also be justified to neglect flavor singlet
contributions (like disconnected insertions and pure gluon operators),
and it makes sense to consider proton and neutron matrix elements separately
(as we have done) and not only flavor non--singlet combinations like
$d_2^{(p)} - d_2^{(n)}$. 
The quark mass dependence of our results is rather mild for the 
range of (relatively large) masses that we studied. Therefore the
extrapolation to the chiral limit looks quite safe, although, of course, 
unexpectedly large effects at truly small masses cannot be excluded.
Lattice artifacts are obvious in our renormalization factors 
(see, e.g., Fig.~\ref{fig.zrat}). We have to expect them also in the 
nucleon matrix elements. Since we are working with non--perturbatively
improved fermions we could in principle reduce their size by using
improved operators. Unfortunately, the non--perturbative improvement
of operators of the kind needed here is not straightforward and has 
yet to be worked out. In particular, improvement of the renormalization
factors requires off--shell improvement. Although there are some ideas on 
how to solve this non--trivial problem (see, e.g., Ref.~\cite{pertpap}),
an implementation for the operators considered here is beyond our present
possibilities. 
 
\section{Conclusions}

In this paper we have tried to obtain a more reliable lattice estimate
of the twist--three nucleon matrix element $d_2$ improving on our 
first calculation~\cite{QCDSF1} in several respects.
We have made a serious
attempt to take into acccount the most important part of the operator mixing
which occurs in this case, namely the mixing with lower--dimensional
operators. This could only be done non--perturbatively and led to a
significant change in the results for $d_2$ moving them close to the 
experimental numbers. Thus the mixing with lower--dimensional operators
seems to account for a large part of the difference between our previous
computation and the experimental data. 

The calculations have been performed in the quenched approximation 
at three different values of $\beta$
corresponding to three different values of the lattice spacing.
While our results are still not good enough to allow for a meaningful
extrapolation to the continuum limit, the mutual consistency of the 
values obtained for $d_2$ at the various $\beta$'s indicates that 
discretization effects are smaller than our statistical errors and
corroborates our 
conclusion that the twist--three nucleon matrix element is rather
small, in agreement with the experimental findings.

We consider the computations at $\beta = 6.4$, i.e.\ at our smallest
lattice spacing, to be most reliable. 
At $Q^2 = 5 \, \mbox{GeV}^2$ they yield the structure function moments 
(cf.\ Table \ref{res5})
\begin{equation}
 \int_0^1 \! \mbox{d}x \, x^2 g_1(x)= \left \{ \begin{array}{rcll}  
  0.017 & \pm & 0.004 & \mbox{(proton)} \\
  0.000 & \pm & 0.002 & \mbox{(neutron)}  \end{array} \right.
\end{equation}
for $g_1$ and 
\begin{equation}
 \int_0^1 \! \mbox{d}x \, x^2 g_2(x)= \left \{ \begin{array}{rcll}  
   - 0.010 & \pm & 0.003 & \mbox{(proton)} \\
   - 0.0002 & \pm & 0.0017 & \mbox{(neutron)}  \end{array} \right.
\end{equation}
for $g_2$. These numbers are to be compared with the
experimental results \cite{abe,anthony}
\begin{equation}
 \int_0^1 \! \mbox{d}x \, x^2 g_1(x)= \left \{ \begin{array}{rcll}  
    0.0124 & \pm & 0.0010 & \mbox{(proton)} \\
   -0.0024 & \pm & 0.0016 & \mbox{(neutron)}  \end{array} \right.
\end{equation}
and 
\begin{equation}
 \int_0^1 \! \mbox{d}x \, x^2 g_2(x)= \left \{ \begin{array}{rcll}  
   - 0.0059 & \pm & 0.0015 & \mbox{(proton)} \\
     0.0029 & \pm & 0.0035 & \mbox{(neutron)}  \end{array} \right.
\end{equation}
respectively.

\section*{Acknowledgment}

The numerical calculations have been done on the APE/Quadrics computers at
DESY (Zeuthen) and the Cray T3E at NIC (J\"ulich) and ZIB (Berlin).
We thank the operating staff for support. This work was
supported in part by the Deutsche Forschungsgemeinschaft and by BMBF.

\begin{table}
\caption{Simulation parameters. In the third column ME indicates the 
calculation of nucleon matrix elements, whereas Z signifies the computation
of renormalization factors. The lattice spacing $a$ has been determined 
from the force scale $r_0$~\protect\cite{sommer} using 
$r_0 = 0.5 \,\mbox{fm}$; 
$c_{\mathrm{SW}}$ is the value of the clover
coefficient. The matrix element calculations for the smallest
quark mass ($\kappa = 0.1353$) at $\beta = 6.4$ have been performed
on a $32^3 \times 64$ lattice.} 
\label{param}
\begin{tabular}{dccdd}
$\beta$ & Lattice & {} & $a^{-1}$[GeV] & $c_{\mathrm{SW}}$ \\ \hline
6.0     & $16^3 \times 32$ & ME & 2.12 & 1.769  \\
6.0     & $24^3 \times 48$ & Z  & 2.12 & 1.769  \\
6.2     & $24^3 \times 48$ & ME & 2.90 & 1.614  \\
6.2     & $24^4$           & Z  & 2.90 & 1.614  \\
6.4     & $32^3 \times 48$ & ME & 3.85 & 1.526  \\
6.4     & $32^3 \times 40$ & Z  & 3.85 & 1.526 
\end{tabular}
\end{table}

\widetext
\begin{table}
\squeezetable
\caption{The unrenormalized reduced matrix elements $a_2$, $d_2^{[5]}$ and
   $d_2^\sigma$ for $u$ and $d$ quarks in the proton at 
   $\beta = 6.0$. Also given are the pion masses.} 
\label{rmeclo60}
\begin{tabular}{cdddddd} 
 {}  & \multicolumn{6}{c}{$\kappa$} \\
   & 0.132 & 0.1324 & 0.1333 & 0.1338 & 0.1342 &
   $\kappa_c$ \\ \hline
$ a m_\pi $  & 0.5412(9) & 0.5042(7) & 0.4122(9) & 0.3549(12) 
             & 0.3012(10) & 0.0 \\ 
 {} & {} & {} & {} & {} & {} & {} \\
$ a_2^{(u)} $  & 0.114(8) & 0.114(8) & 0.107(11) & 0.09(2)
               & 0.08(2) & 0.08(2) \\
$ a_2^{(d)} $  & $-$0.029(3) & $-$0.032(4) & $-$0.037(7) & $-$0.032(12) 
               & $-$0.047(16) & $-$0.046(11) \\ 
 {} & {} & {} & {} & {} & {} & {} \\
$ d_2^{[5]\,(u)} $  & 0.0063(12) & 0.0028(13) & $-$0.010(2) 
          & $-$0.023(5) & $-$0.028(6) & $-$0.037(4) \\ 
$ d_2^{[5]\,(d)} $  & $-$0.0041(6) & $-$0.0027(7) & $-$0.0008(14) 
          & $-$0.005(2) & $-$0.001(4) & 0.001(2) \\ 
 {} & {} & {} & {} & {} & {} & {} \\
$ d_2^{\sigma\,(u)}/a $  & $-$0.216(12) & $-$0.228(12) & $-$0.246(18) 
          & $-$0.29(4) & $-$0.27(3) & $-$0.30(3) \\ 
$ d_2^{\sigma\,(d)}/a $  & 0.050(4) & 0.050(3) & 0.052(6) 
          & 0.064(12) & 0.044(17) & 0.057(12) 
\end{tabular}
\end{table}

\begin{table}
\squeezetable
\caption{The unrenormalized reduced matrix elements $a_2$, $d_2^{[5]}$ and
   $d_2^\sigma$ for $u$ and $d$ quarks in the proton at 
   $\beta = 6.2$. Also given are the pion masses.} 
\label{rmeclo62}
\begin{tabular}{cddddd} 
 {} & \multicolumn{5}{c}{$\kappa$} \\
   & 0.1333 & 0.1339 & 0.1344 & 0.1349 & 
    $\kappa_c$ \\ \hline
$ a m_\pi $  & 0.4136(6) & 0.3570(10) & 0.3034(6) & 0.2431(7) & 0.0 \\
 {} & {} & {} & {} & {} & {} \\
$ a_2^{(u)} $  & 0.142(10) & 0.137(15) & 0.157(17) & 0.16(3) & 0.17(3) \\ 
$ a_2^{(d)} $  & $-$0.033(4) & $-$0.030(5) & $-$0.034(9) & $-$0.031(15) 
               & $-$0.030(13) \\  
 {} & {} & {} & {} & {} & {} \\
$ d_2^{[5]\,(u)} $  & 
     $-$0.0017(14) & $-$0.017(2) & $-$0.031(5) & $-$0.051(11) & $-$0.065(7) \\ 
$ d_2^{[5]\,(d)} $  & $-$0.0027(6) & 0.0006(11) & $-$0.0003(17) & $-$0.000(4) 
                    & 0.004(3) \\ 
 {} & {} & {} & {} & {} & {} \\
$ d_2^{\sigma\,(u)}/a $  & 
     $-$0.34(2) & $-$0.37(3) & $-$0.44(4) & $-$0.50(7) & $-$0.54(7) \\ 
$ d_2^{\sigma\,(d)}/a $  & 0.067(6) & 0.065(8) & 0.072(13) & 0.07(3) & 0.07(2) 
\end{tabular}
\end{table}

\begin{table}
\squeezetable
\caption{The unrenormalized reduced matrix elements $a_2$, $d_2^{[5]}$ and
   $d_2^\sigma$ for $u$ and $d$ quarks in the proton at 
   $\beta = 6.4$. Also given are the pion masses.}
\label{rmeclo64}
\begin{tabular}{cdddddd} 
 {}  & \multicolumn{6}{c}{$\kappa$} \\
   & 0.1338 & 0.1342 & 0.1346 & 0.135 & 0.1353 &
  $\kappa_c$ \\ \hline
$ a m_\pi $  & 0.3213(8) & 0.2836(9) & 0.2402(8) & 0.1933(7) 
             & 0.1507(8) & 0.0 \\ 
 {} & {} & {} & {} & {} & {} & {} \\
$ a_2^{(u)} $  & 0.123(8) & 0.092(13) & 0.114(13) & 0.102(17)
               & 0.14(4) & 0.096(19) \\
$ a_2^{(d)} $  & $-$0.030(4) & $-$0.018(5) & $-$0.032(7) & $-$0.033(10) 
               & $-$0.01(2) & $-$0.023(10) \\ 
 {} & {} & {} & {} & {} & {} & {} \\
$ d_2^{[5]\,(u)} $  & $-$0.0149(14) & $-$0.024(3) & $-$0.038(4) 
          & $-$0.055(6) & $-$0.055(15) & $-$0.068(6) \\ 
$ d_2^{[5]\,(d)} $  & $ 0$.0014(7) & $ 0$.0021(13) & $ 0$.0062(17) 
          & $ 0$.012(3) & $ 0$.015(8) & 0.013(3) \\ 
 {} & {} & {} & {} & {} & {} & {} \\
$ d_2^{\sigma\,(u)}/a $  & $-$0.40(2) & $-$0.41(4) & $-$0.47(4) 
          & $-$0.51(5) & $-$0.55(13) & $-$0.56(5) \\ 
$ d_2^{\sigma\,(d)}/a $  & 0.088(6) & 0.075(12) & 0.105(13) 
          & 0.12(2) & 0.15(5) & 0.12(2) 
\end{tabular}
\end{table}

\begin{table}
\caption{Results for $\mu^2 = Q^2 = 5 \, \mbox{GeV}^2$.}
\label{res5}
\begin{tabular}{cddd} 
 {}  & \multicolumn{3}{c}{$\beta$} \\
   & 6.0 & 6.2 & 6.4 \\ \hline
$ a_2^{(p)} $ & 0.046(13) & 0.11(2)  & 0.066(14) \\
$ d_2^{(p)} $ & 0.008(4)  & 0.017(9) & 0.017(7)  \\
$ \int_0^1 \mbox{d}x x^2 g_1^{(p)}(x) $
              & 0.012(3) & 0.029(5) & 0.017(4) \\
$ \int_0^1 \mbox{d}x x^2 g_2^{(p)}(x) $
              & $-$0.007(3) & $-$0.019(4) & $-$0.010(3) \\
 {} & {} & {} & {} \\
$ a_2^{(n)} $ & $-$0.017(8) & 0.009(11)    & 0.000(8) \\
$ d_2^{(n)} $ & $-$0.003(2) & $-$0.001(4)  & $-$0.001(3) \\
$ \int_0^1 \mbox{d}x x^2 g_1^{(n)}(x) $
              & $-$0.004(2) & 0.002(3) & 0.000(2) \\
$ \int_0^1 \mbox{d}x x^2 g_2^{(n)}(x) $
              & 0.0026(16) & $-$0.002(2) & $-$0.0002(17)
\end{tabular}
\end{table}

\begin{table}
\caption{Results for $\mu^2 = Q^2 = 10 \, \mbox{GeV}^2$.}
\label{res10}
\begin{tabular}{cddd} 
 {}  & \multicolumn{3}{c}{$\beta$} \\
   & 6.0 & 6.2 & 6.4 \\ \hline
$ a_2^{(p)} $ & 0.040(12) & 0.098(18) & 0.057(12) \\
$ d_2^{(p)} $ & 0.002(4)  & 0.006(7)  & 0.008(6)  \\
$ \int_0^1 \mbox{d}x x^2 g_1^{(p)}(x) $
              & 0.010(3) & 0.025(5) & 0.015(3) \\
$ \int_0^1 \mbox{d}x x^2 g_2^{(p)}(x) $
              & $-$0.007(2) & $-$0.017(4) & $-$0.009(3) \\
 {} & {} & {} & {} \\
$ a_2^{(n)} $ & $-$0.015(7) & 0.007(9) & 0.000(7) \\
$ d_2^{(n)} $ & $-$0.003(2) & $-$0.002(3)  & $-$0.001(3) \\
$ \int_0^1 \mbox{d}x x^2 g_1^{(n)}(x) $
              & $-$0.0037(18) & 0.002(2) & 0.0001(19) \\
$ \int_0^1 \mbox{d}x x^2 g_2^{(n)}(x) $
              & 0.0021(14) & $-$0.0016(18) & $-$0.0002(14)
\end{tabular}
\end{table}

\end{document}